\documentclass{article}
\usepackage{ijcai21}

\usepackage{times}

\usepackage{soul}
\usepackage{url}
\usepackage[hidelinks]{hyperref}
\usepackage[utf8]{inputenc}
\usepackage[small]{caption}
\usepackage{graphicx}
\usepackage{amsmath}
\usepackage{booktabs}
\usepackage{color, colortbl}

\usepackage{makecell} 
\usepackage{tabularx}

\usepackage[hidelinks]{hyperref}
\usepackage[T1]{fontenc}

\definecolor{LightCyan}{rgb}{0.88,1,1}
\definecolor{LightGrey}{gray}{0.9}
\definecolor{Purple}{rgb}{.6,0,.6}
\definecolor{ltgray}{gray}{0.9}

\newcommand\mmm[1]{\textcolor{black}{#1}}
\newcommand\kjf[1]{\textcolor{black}{#1}}
\newcommand\nnn[1]{\textcolor{black}{#1}}

\makeatletter
\newcommand*\titleheader[1]{\gdef\@titleheader{#1}}
\AtBeginDocument{%
	\let\st@red@title\@title
	\def\@title{%
		\bgroup\normalfont\large\centering\@titleheader\par\egroup
		\vskip1.5em\st@red@title}
}
\makeatother

\title{Informing Autonomous Deception Systems with Cyber Expert Performance Data}
\titleheader{\emph{IJCAI-21 1\textsuperscript{st} International Workshop on Adaptive Cyber Defense}}
\author{
Maxine M. Major$^1$\and
Brian J. Souza$^{1}$\and
Joseph DiVita$^1$\And
Kimberly J. Ferguson-Walter$^2$\\
\affiliations
$^1$Naval Information Warfare Center Pacific\\
$^2$Laboratory for Advanced Cybersecurity Research\\
\emails
\{mmajor, bsouza\}@niwc.navy.mil, joseph.divita@navy.mil, kimberly.j.ferguson-walter.civ@mail.mil
}

\begin{document}

\maketitle

\begin{abstract}
The performance of artificial intelligence (AI) algorithms in practice depends on the realism and correctness of the data, models, and feedback (labels or rewards) provided to the algorithm. This paper discusses methods for improving the realism and ecological validity of AI used for autonomous cyber defense by exploring the potential to use Inverse Reinforcement Learning (IRL) to gain insight into attacker actions, utilities of those actions, and ultimately decision points which cyber deception could thwart. The Tularosa study, as one example, provides experimental data of real-world techniques and tools commonly used by attackers, from which core data vectors can be leveraged to inform an autonomous cyber defense system. 
\end{abstract}

\section{Introduction}~\label{sec:intro}
One of the most pervasive problems in technology today is that given enough time and resources, a cyber attacker nearly always has the advantage over a defender. If persistent, they will ultimately unveil a weakness that can then be exploited. The difficulty of maintaining complex networks has historically constrained defenders to be truthful about the topology and configuration of networks, which \kjf{has had} unintended consequences that benefit malicious actors. However, given the option of using cyber deception, a defender can level the playing field by casting doubt on the attacker's understanding of the networks they wish to attack. The concept of adding deception to cyber defense has been discussed for decades~\cite{tinnel2002}\cite{yuill_2006} \cite{michael_phase_2004}, with recent experiments proving the effectiveness of various deception techniques~\cite{usenix-2021}\cite{10.1007/978-3-030-20488-4_11}\cite{moonraker-2020}. However, with the addition of this new layer of detection and mitigation techniques come a new set of challenges. With an established deception ruleset, cyber adversaries may learn what kinds of deception to expect, and in turn learn to avoid it. As attackers continuously discover ways to work around or avoid the deceptive defenses, cyber defenders must in turn adapt and change the defenses in continuous attempts to thwart attacks.

An autonomous cyber deception solution can ease the burden on defenders and make intelligent, timely decisions on their behalf. A real-time, adaptive deception system, which can ingest inputs from defensive and deceptive sensors, can learn from adversarial activity in order to track and respond with customized deception to deter, delay, or misinform potential attackers~\cite{fugate2019}. Based on the defender's own terrain, this system would naturally be limited to data that is available to the defender, either through their own Intrusion Detection System (IDS) or security systems, or through deception tools such as decoys or tripwires planted in the network. Additionally, prior research and statistical analysis about attacker behavior --- including how attack patterns change in the presence of deception --- can inform such models. 

A major component driving the decisions for the type and placement of deception is to learn the \textit{reward} for taking a particular deceptive action, which in turn is based on the \textit{utility} of that action - which requires a predictive look-ahead at the probabilities of the attacker taking a particular course of action. This requires an intelligent cognition of the current state of defense, awareness of which data have been discovered and interacted with by an attacker, and what the attacker's own likely action rewards and utilities might drive them to do next. 
This paper explores the data available to a defender to learn the current state of attack and to estimate the possible rewards and utilities of both the attacker and a deceptive response, with the goal of using this information to inform an autonomous cyber deception defense. The paper draws on observations from the Tularosa study where red teamer behavior was compared in a large scale cyber deception experiment~\cite{tularosa2019}. Network and host actions were recorded and behaviors and beliefs about their own performance and deception were self-reported. We discuss how insights from data of this type can be used for modeling attacker and defender rewards for autonomous defense systems.

\section{Background \& Related Work}~\label{sec:related_work}
The concept of using deception to gain an advantage over the enemy is not a new one---deception has been employed throughout history to great wartime advantage. Modern cyber deception can be used for early threat detection, and to understand and subvert the cyber enemy. Advances in deception techniques have demonstrated that cyber deception is applicable across all attack surfaces~\cite{dickinson2020}. Along with advancements made in automating cyber defense, a real-time, autonomous deception system to respond to an unseen adversary's advances is a natural technological progression. 

\subsection{Autonomous Defenses}
\nnn{Cyber defense is a multi-faceted role, and a human defender is limited in their ability to monitor all threat vectors simultaneously. As a result, \kjf{the goal of increased} automation is quite prevalent in cyber defense technologies. Autonomous cyber defense \kjf{also requires automating decision making, which is well-suited for artificial intelligence (AI) solutions}---reinforcement learning (RL) in particular, can address many of the challenges. Using RL to select various cyber defense policies has been shown to improve cyber risk, compared to a random selection of valid actions~\cite{beaudoin2009}. Intelligent automation can also learn to detect and automatically patch cyber threats, as demonstrated by several of the final round solutions in the DARPA Cyber Grand Challenge (CGC)~\cite{darpa2015} and RL scenarios defending against a randomized adversary~\cite{ridley2018}. \kjf{Furthermore, it is clear that} adaptive automation which can keep up with changes in technology and attack strategies is ideal for the ever-evolving technological landscape~\cite{crawford2017}. }

\subsection{Deception Technologies}
Deception is becoming a mainstream feature for many commercial cyber defense tools, along with their standard IDS feature set. One strategy is to scan the existing network topology, including users, services, and assets, to design decoys or decoy subnets which mimic the features of that network~\cite{fidelis2021}~\cite{cybertrap2021}~\cite{smokescreen2021}. Due to the proprietary nature of commercial technologies, comparing the robustness of the underlying deception technologies is difficult. Much work remains to be done to make these tools truly autonomous; While a few claim to automatically adapt their deception strategy based on observed malicious activity, many of these act based on preset rules or have limited adaptability, and several still use default deception configurations or require a defender's hands-on assistance to implement new deception strategies~\cite{trapxsecurity2017}. Advancements in optimizing the type and placement of cyber deception, along with removing the burden of management from the defender, will improve the robustness of cyber deception platforms and the state of cyber defense overall.

\subsection{Defensive Deception Research}

Deception in a cyber context aims to disrupt cyber attacks by presenting false information to an attacker, which they can believe or act upon. Deception encompasses more than simple honeynets and virtual machines meant to distract attackers; they can be classified into four categories; \textit{conceal facts}, \textit{reveal fictions}, \textit{reveal facts}, and \textit{conceal fictions}, as per the Cyber Denial \& Deception (D\&D) Matrix~\cite{heckman_cyber_2015}.

Ongoing research aims to produce new methods for detecting and mitigating cyber threats through dynamic, intelligent cyber deception, based on attacker activity. IDS alerts have been used in order to launch pre-configured honeynets~\cite{acosta2020}.
On a host-level, machine-learning and malware behavioral analysis techniques can turn attackers' own malware deception back onto them, altering the attacker's decision making~\cite{sajidul2020}. A prototype called KAGE used software defined networking (SDN) to establish an ``alternate reality'' among distributed systems, deploying deceptive counterparts to attacker-desired services in real time, based on pattern-matching rules~\cite{soule2016}. 

Game theory has often been used to model the interaction between the cyber attacker and defender and investigate the effect of deception on attacker performance~\cite{ferguson-walter_gamesec_2019}\cite{carroll2009}\cite{feng2017}\cite{kovach2015}. Human subjects research (HSR) can both complement and expand the scope of cyber game theory by including the cognitive decision-making biases of the attacker~\cite{gutzwiller2019}\cite{johnson2020}. 

Notable research evaluated different non-deceptive defense techniques against an active attacker in a cyber security simulation modeled as a Markov game with incomplete information. Techniques evaluated include Monte Carlo and Q-Learning, with the best performance from those able to adapt to the changing attack and defense scenario \cite{elderman2017}. Cyber deception strategies similarly need to adapt to changes in pace with the attack strategy. Attacker profiles developed via an attacker taxonomy which inferred their goals and skill level, were pitted against simulated cyber deception to evaluate the difference in how attackers of varying skill level are affected by deception~\cite{8328465}. 

Although automation seems ideal for effective cyber defense including defensive deception, we note that this could lead to an over-reliance on the system by a human defender, thus weakening the effectiveness of the defender, and other "ironies of automation"~\cite{gutzwiller2021}. Further research will need to examine how to maximize the benefits for, and minimize the overhead on human defenders.

\subsection{The Human Attacker}
\nnn{Considering deception to be primarily a \mmm{cyber defense technology}} helps to scope the problem space, but \nnn{is limiting} due to the nature of the cyberpsychology behind deception. \nnn{Deception has an impact on the target's cognitive state, and thus defensive cyber deception has the potential to learn about \kjf{and impact} the decision-making human behind the attack, in order to to improve predictiveness and effectiveness.} Behavioral psychology has demonstrated the ability to completely flip the outcome of risk-aversion versus risk-seeking, by manipulating the context, or frame, under which the decision is made. \nnn{Because of this}, human decision-making may be manipulated by controlling the context of the decision~\cite{kahneman1979}\cite{tversky1974}. 

If a cyber defender can frame a decision that the attacker must make, using behavioral science research to predict with great likelihood the outcome of the attacker's decision, then the defender may anticipate the attacker's next move and gain an exploitable advantage. 
The most meaningful actions an attacker takes toward their goal likely occur on the defender-owned domain, and these actions progressively build upon prior work performed toward those goals. Each action in the cyber attack can be decomposed into incremental actions within stages of a cyber attack, such as the stages of the Cyber Kill Chain~\cite{lockheed}. By explicitly discovering the decisions, and actions relevant to each detectable stage of an attack, deception may be used to impact that decision. 
Attacker motivation is a difficult problem to solve. In behavioral research, decisions under conditions of uncertainty are posed in terms of the probabilities of receiving a reward or penalty~\cite{kahneman1979}. 
In the context of a cyber attack, the reward would refer to the perceived value of the target to the attacker. The probability of receiving this reward corresponds to the probability of success the attacker associates in carrying out the attack. This comparison suggests two immediate research questions (both of which are impacted by deception): 
\begin{enumerate}
	\item How does an attacker determine the value of a target? I.e., what attributes and signals from a target do attackers use to determine the value of attacking that target?
	\item How do attackers evaluate the probability of launching a successful attack? I.e., what attributes and signals from a target do attackers use to determine the vulnerabilities and likelihood of successfully compromising a target? 
\end{enumerate}

\nnn{By learning these attributes and features}, experimentation may be designed to test their effect on the attacker's decision making, and to predict whether the attacker's decisions exhibit risk-seeking or risk-averse behavior. \nnn{For} attacker behavior that conforms to these predictions, \nnn{deception can be deployed to} bait the attacker into making a series of decisions that the defender anticipates and \mmm{further} exploits. 
\nnn{Effective deception can be used to manipulate} an attacker's assessment of a target's value, the attacker's perception of the likelihood of achieving their goals against that target, \nnn{and then} strategically incorporate \nnn{these strategies into cyber defense} systems.

 While most adaptive systems utilizing AI require the ability to reason under uncertainty, when utilizing deception in an adversarial scenario, the system will likely need to represent the attacker's incorrect perception of the network that the defender's deception is providing to the attacker. \nnn{This can} be modeled as a hypergame~\cite{ferguson-walter_hotsos_2019}. 

\nnn{Information about attackers' internal motivations and beliefs is elusive to capture in the wild. However, experimentation \kjf{providing cyber expert performance data,} such as the Tularosa Study~\cite{tularosa2019}, can provide a rare glimpse into the cognition of hacker-like populations: professional red-teamers and penetration testers, which is discussed in more detail in the following Section~\ref{sec:tularosa}.}

%
\subsection{The Tularosa Study}~\label{sec:tularosa}

The Tularosa Study was a controlled human-subjects research (HSR) experiment \nnn{which aimed to gain insight into the human side of attacker strategies and beliefs. For two days,} professional red team hackers were given a cyber hacking exercise on identical copies of a simulated network~\cite{tularosa2019}. Participants were either presented a network with decoys (deception-present condition), were only~\textit{told} that there might be deception (psychological condition, \mmm{"reveal fictions"}), both (\mmm{"conceal \& reveal fictions"}), or neither (for the control condition). Data collected from this experiment includes cyber data from sources that would only be visible to the attacker (keylog, screen recordings), data usually accessible to the defender (network traffic, IDS alerts), and data only available when deception is present (decoy alerts). Additional data which would not be available in the wild, but can help characterize attacker behavior and improve defenses includes: cognitive tests, self-reports, and demographics data. \nnn{Results from the cognitive and self-report data helps demonstrate deception's impact to both cyber and psychological performance, reinforces the utility of defensive deception, and illustrates deception's impact on attacker thoughts and behaviors.}

Data analysis reveals that deception --- both cyber and psychological --- impacts the attacker's actual ability to make progress toward compromising real targets~\cite{usenix-2021}. Cognitive data analysis reveals that red teamers who believed in deception were more likely to express confusion, self-doubt, and become frustrated during this task~\cite{fergusonwalter-thesis}. The information from this experiment supports the development of a rewards model and autonomous deception solution via two main contributions: 1) provide context for how defender data can inform deception responses and provide feedback with respect to the effectiveness of those responses on an attacker, and 2) inform the development of an attacker model and implicitly measure the attacker's response to that deception.

Throughout this paper, references will be made to data and insights gleaned from the Tularosa Study---as an example--demonstrating how observations gained from experiments can support \nnn{the understanding of how deception impacts human attacker cognition, supports the} development of a \kjf{realistic} reward function, and assists \nnn{in developing attacker profiles}. Further details on the study can be found in previous publications~\cite{ferguson-walter_HACS_2019}\cite{tularosa2019}\cite{tularosa_appendix}.

\nnn{Data points collected from the defensive, deceptive environment which can be used to detect attackers and monitor for deception's effectiveness are discussed further in Section~\ref{sec:data}. A method which can be used to infer attacker decision points and beliefs, and \kjf{derive a reward function} which can help drive autonomous deception solutions, is Inverse Reinforcement Learning (IRL), detailed in Section~\ref{sub:irl}.}

\section{Data for Adaptive Cyber Deception}~\label{sec:data}
\mmm{In the} real-world, cyber defenders \mmm{typically} do not have visibility into the attacker's goals, strategies, beliefs, skills and individual techniques leveraged to accomplish their nefarious deeds. \mmm{Early stages of attack---initial reconnaissance in particular---tend to be quiet; the defender often learns about a skilled, successful attacker during the later stages of an attack, if at all. Even \nnn{after} an attack has fully succeeded, the defender can only make guesses about the attacker's true goals and skills by piecing together artifacts from the attacker's tracks after the fact. This is a result of extremely limited} data available \mmm{for a defender to} detect malicious actors or evaluate the results of deception \kjf{using only} data gleaned from their own network and resources. 

\mmm{Cyber deception provides a unique advantage for defenders to gain early alerting of cyber attacks through simple methods such as canaries and other tripwires, \kjf{and information on attacker preferences and behaviors through honeypots, honeytokens, and decoys}. There also exists a psychological component by which the defender gains the potential to direct the attacker's own actions 
by leveraging intelligent, adaptive deception to alter the attacker's beliefs about the network and their own success.}
\mmm{Standard} defender-available \mmm{network and host} data can \mmm{directly support the model and decisions made by} an intelligent deception response tool.

The following sections discuss types of \mmm{standard} information available to a cyber defender, which, along with knowledge of \mmm{known} attacker attributes, \mmm{can model} a feature space of \mmm{networked} resources \mmm{(targets)}, actions available to both attackers and deceptive defenders, a model of the state space, and aspects of creating a custom profile for the attacker.

\subsection{Feature Space}~\label{sub:feature}
With full visibility into network traffic activity, and limited visibility into host-based activity, along with certain types of deception, such as decoys and false network packets, defenders can create a custom feature space to gather attacker activity and respond with deception. The feature space for modeling a deceptive response consists of practical information that is readily available to the defender, \nnn{and with} which the attacker is likely to interact with or use in executing their strategies. This information includes data about the network and hosts, but also includes activities detected on the network and how deception or adversaries might alter properties of these features. \mmm{Ultimately, the goal is for an autonomous defensive deception capability to implement and optimize the deception's effectiveness by continuously evaluating malicious interactions on these resources.}

We present a small subset of the possible features in order to highlight the process in which experimental results can be used to inform intelligent defensive systems, to ensure the learning is based on models and rewards that are grounded in reality. For the simple rewards model presented in this paper, we consider features in Table~\ref{table:features}. This \mmm{highly-simplified} collection of network and host knowledge \mmm{only represents} a very narrow subset of thousands of unique features available from any given host which may \mmm{be impacted by} attacker activity. 

\bgroup
\def\arraystretch{1.5}
\begin{table}[!htb] \small
	\begin{tabular}{ | p{0.35\columnwidth} | p{0.55\columnwidth} | }
		\hline
		\small
		\cellcolor{ltgray} Feature Category & \cellcolor{ltgray}Feature \\ \hline
		Domain host properties & \texttt{IP\_address, OS, is\_decoy, services, OS\_ver, svc\_versions, user\_account}\\ \hline
		Domain accounts & \texttt{user\_accounts, is\_decoy, priv\_level, password, pwd\_hash} \\ \hline
		Interconnectivity & \texttt{node\_IP, conn\_type} \\ \hline
		Vulnerabilities & \texttt{service}, \texttt{svc\_version}, \texttt{svc\_vuln}  \\ \hline
		Host metadata & \texttt{purpose}, \texttt{value}  \\ \hline
		Network Activity	& \texttt{traffic\_alert}, \texttt{IDS\_alert} \\ \hline
		Host Activity	& \texttt{HBSS\_alert}, \texttt{local\_alert} \\ \hline
		Metadata	& \texttt{num\_hosts, num\_decoys, num\_adversaries, time\_since\_alert} \\ \hline
	\end{tabular}
	\caption{\textbf{Features:} A simplified set of characteristics of the network and hosts which a deceptive defender can monitor for changes, or \mmm{about} which to \mmm{implement} deception. Features also include metadata \mmm{to model} the current attack state.} 
	\label{table:features}
\end{table}
\egroup

Each time an adversary interacts with or modifies one of these resources, or when a deception is applied to that resource, these changes will be logged as part of that particular state. 
\nnn{At a minimum, each of these features can be treated as a tripwire, while nuanced changes and interactions with these features can track a target's interest, strategies, and overall response to deception.} 
\mmm{For example, often attackers who gain a foothold on a target will attempt to do so through leveraging an existing user account, ideally one with administrator privileges, or attempt to escalate privileges to administrator level.} In the Tularosa experiment, \mmm{several participants attempted to use an existing domain admin account, however,} participants with deception present---who were also informed of deception---attempted to use a compromised domain admin account \textit{less} than participants in the control condition~\cite{usenix-2021}. \mmm{Not only does this demonstrate that attackers may alter or dampen their attack in the presence of deception, but that knowing about the deception may cause reduced or altered interactions \nnn{with these features}.} 
	
\mmm{Data on} the patterns of domain user account usage \mmm{is already provided via defender-owned resources, which can also help a cyber deception capability} track the effectiveness of the current deception response. \mmm{The} host information is simple in this example, but other \mmm{more nuanced} host-based evidence of malicious activity such as tampering with system files or the start of novel processes can also reveal the presence of an attacker, and is the focus for much ongoing research toward novel threat detection.

Cyber defenders \mmm{normally} leverage network and host-based monitoring to detect and observe adversarial activity. \mmm{Common tools such as} an IDS \mmm{are designed to} generate alerts for suspicious network traffic, however, tracking the frequency and severity of these alerts can also \nnn{reveal patterns about} adversarial behaviors. 
\nnn{As an example,} Tularosa Study participants with deception present sent less bytes to real targets, and less packets overall. The popular EternalBlue exploit yielded success on several real targets\nnn{, however,} Tularosa participants whose EternalBlue exploit attempts on real targets were detected by the Suricata IDS yielded twice as many alerts in the deception-absent condition as those with deception-present, and the least number of alerts were triggered by participants who were also informed of deception. 

\nnn{By tracking these fundamental features of both real and deceptive resources, and through careful manipulation of the properties of these resources, deception can provide not only early warning of adversarial interactions, but overall patterns of behavior and changes in response to both real and deceptive features on the network.}

\subsection{Action Space}~\label{sub:action}

\mmm{Cyber defenders and attackers have diametrically opposed goals, and this is reflected in the diversity of actions and strategies utilized by each. Attackers aim to discover, exploit, and claim ownership while remaining undetected; defenders want to protect resources, preserve ownership, prevent exploitation, and reduce the threats posed by attackers when possible. Deception includes the added goals of delaying, deterring, and polluting the knowledge of the attacker.}
Each of \mmm{the attacker's and defender's} actions \mmm{can be considered} \nnn{to occur} in relation to the feature types listed in Table~\ref{table:features}, \mmm{along with \nnn{timestamped} metadata such as} properties added, modified, or changed. \mmm{Evaluating a} history of these actions over time helps to flesh out implicit values for those features\mmm{, learn the most effective deception response,} and \mmm{curate a profile of the attacker}, as depicted in Section~\ref{sub:profile}. \mmm{While it may be difficult to definitively learn the intrinsic motivations and skills of an attacker, the best place to start is by using data that can reveal evidence of the attacker's actions, and through that data, \nnn{predict} the attacker's \nnn{preferences during an}  attack.}

\subsubsection{Attacker Action Space}\label{sub:attacker_actions}
The actions of an attacker are loosely inspired by the commonly adopted, high-level cyber attack sequence model, the Cyber Kill Chain~\cite{hutchins_intelligence-driven_2011}. The actions depicted in Table~\ref{table:attacker_actions} describe categories of activity that an attacker may take, which roughly correspond to the order of actions taken in the Cyber Kill Chain, but are simplified for the sake of focusing on activities to which a defender may discover and respond.

\bgroup
\def\arraystretch{1.5}
\begin{table}[!htb]\small
	\begin{tabular}{ | p{0.35\columnwidth} | p{0.55\columnwidth} | }
		\hline \small
		\cellcolor{ltgray} Adversarial Action & \cellcolor{ltgray}Description\\ \hline
		\texttt{do\_nothing} & Null or ``'wait``. The state has not changed, or no action recommended.\\ \hline
		\texttt{\thead[l]{passive\_recon}} & Observe network traffic, identify network hosts and host roles. Not detected unless HBSS monitoring for this activity resides on the host being used for passive recon. \\ \hline 
		\texttt{\thead[l]{active\_recon}}& Probe individual hosts to learn specific features, e.g., OS, ports, services, and service versions. \\ \hline
		\texttt{\thead[l]{vuln\_search}}& Search for weaknesses on known services. Can be performed ``offline'' on the attacker's own machine, or via a vulnerability scan, which is visible by the defender. \\ \hline
		\texttt{\thead[l]{explore\_service}} & Discover information not included in scans which does not require an exploit or login, e.g., web pages, or unauthenticated services (e.g. telnet). \\ \hline
		\texttt{exploit} & Exploit a vulnerability to allow entry or obtain advanced permissions.\\ \hline
		\texttt{\thead[l]{actions\_target}} & Malicious activity after foothold obtained, e.g., exfiltration, backdoors, sabotaging resources. \\ \hline
	\end{tabular}
	\caption{\textbf{Adversarial Action Space:} Simplified list of high level adversarial actions.} 
	\label{table:attacker_actions}
\end{table}
\egroup

Although theoretically all actions in Table~\ref{table:attacker_actions} are available to an attacker with a foothold, not all actions are equally valuable for the state the attacker is in. For example, actions which provide information, e.g., \texttt{active\_recon}, should be of higher value than \texttt{exploit}, if launching an exploit from the current state is likely to fail if vulnerabilities are not known. If an attacker has not yet scanned the network, then \texttt{passive\_recon} would have more utility than \texttt{vuln\_search} on targets whose identities are unknown. 
However, despite the fact that these actions seem logically taken in an orderly sequence, once the attacker has gained minimal information from the target domain, even as early as the \texttt{passive\_recon} stage, depending on the attacker's goals, strategies, and even skill set, the utility of taking any one action on any number of possible targets can make it difficult to predict the \nnn{likely next actions taken, and thus the} best deception to mitigate these actions. 
\mmm{Toward predictive modeling of an attacker's likely next move, based on prior knowledge of attacker actions, an} autonomous deceptive agent could be ideal for tracking and responding \mmm{predictively to these} patterns of activity.

While our description is based on the premise that every attacker activity will be detected by defenders, this is purely for the purpose of demonstrating that deception \nnn{can be effective against} every attacker action, \mmm{detectable or not}. Highly skilled cyber actors often \mmm{avoid detection by acting under the guise of} normal traffic and user activity. However, as we see in Table~\ref{table:defender_actions}, several types of deception can serve as traps to alert the defender to the type of \mmm{early stage or covert} adversarial activity that would normally go undetected. 
\mmm{In order to identify the types of deception that are appropriate to deploy against potential attackers, a defender should consider the types of resources in their domain, and plan counter-deception strategies against the same attack tools and techniques that are commonly used against those types of targets. Automatic network discovery tools, commonly found bundled with commercial IDS products, could partner with detailed attack taxonomies to reduce overhead for the defender. For example,} the ATT\&CK~\cite{mitre_attck} matrix outlines very specific tools and techniques that could be taken by an attacker, and categorizes them along an expanded attack chain which includes details about tools or exploits that may be used, any files altered, and residual effects in the target system. 
ATT\&CK is leveraged by tools including Atomic Red Team~\cite{atomicredteam}, and CALDERA\texttrademark~\cite{mitre_caldera} to emulate and automate a given cyber actor or campaign. These, and similar resources, can support additional research to \mmm{learn and} define the attack space, the feature space for each attack \mmm{technique}, and \mmm{ultimately} support \nnn{deceptive responses} targeting these \nnn{actions}.

\subsubsection{Defensive Deception Action Space}\label{sub:defender_actions}

The actions available to a deceptive defender fall roughly into one of three categories: 1) actions typically available to the cyber \textit{defender}'s role, which pertain to normal day-to-day activity, such as monitoring and maintenance, 2) non-deceptive responses to suspicious or malicious activity, \nnn{such as blacklisting malicious IPs}, and 3) actions specific to \textit{deception}, which complement defenses by adding deceptive features and strategies. Rather than enumerating the types of defensive actions\mmm{, or classifying as per the Cyber D\&D Matrix~\cite{heckman_cyber_2015},}  here, we \mmm{consider a simplified list of} data-collecting activities which support deception-based actions in Table~\ref{table:defender_actions}. For a list of non-deceptive active responses we direct readers to consult the Open Command and Control (OpenC2) Language Specification~\cite{openC2} which provides 32 actions, none of which are deception-specific.

The deception action space is where the potential to really disrupt \nnn{and learn from} a cyber attacker comes into play~\cite{fugate2019}. 
\nnn{Deception does not need to be complicated;} the Tularosa Study was limited to static, low-interaction decoys. During scans the \nnn{decoys} appeared to be legitimate Windows or Linux hosts with a few believable services running. This created delays for attackers who wasted time investigating, searching for vulnerabilities, \kjf{or attempting to exploit} these hosts. \mmm{By definition, \textit{any} time spent interacting with decoys is time wasted, as the presence and properties of these decoys provided no benefit toward compromising real hosts on the network.} Although the red teamers eventually moved on to find other targets that responded more \kjf{predictably}, the result of wasting time on these false targets resulted in less IDS alerts, less attempts to use a domain admin account, less exfiltrated files, and less success on real hosts overall~\cite{usenix-2021}. From the perspective of wasted time alone, static, low-interaction decoys were more \kjf{effective} than no deception at all. 

Table~\ref{table:defender_actions} includes a list of potential deception actions which can be used for \nnn{either} initial deception or \nnn{dynamically modified for continuous disruption of} a given attacker's goals. 
\nnn{Several of the deception actions itemized in Table~\ref{table:defender_actions} address multiple attacker actions from Table~\ref{table:attacker_actions}, since a single type of deception, particularly those early in the Kill Chain, has the potential for lingering effects throughout the attack lifecyle. For example, a \texttt{ping\_responder} in place, can serve as an early warning for quiet reconnaissance techniques while convincing the attacker that ``something'' resides at that network location. With this early response, deceptive tools such as \texttt{traffic\_control} or \texttt{TCP\_reset} can slow or drop network scans, allowing for \texttt{launch\_decoy} to plant false targets deeper into the subnet before scanning can complete, or to even create false subnets (or a sandbox) customized with similar targets to those the attacker is known to probe. \kjf{The actions listed are the highest-level description of actions that will typical have a hierarchical structure. For example once the \texttt{launch\_decoy} action is selected by the AI, the configuration of the decoy will also need to be determined such as the OS, patch level, open ports, and services running. Details from the how the participants responded to particular decoys in the Tularosa Study, for example, can be used to build initial attacker profiles. This can help determine which decoy configuration is optimal given the defender's goal in relation to manipulation of the attacker.} }
	
	\nnn{With tools such as \texttt{portspoof}~\cite{portspoof}, the properties of both real and decoy targets can be falsely presented, which disrupts the attacker's \texttt{active\_recon}, \texttt{vuln\_search}, and \texttt{exploit} activity. High interaction decoys can disrupt the attacker's progress further, with working exploits and a fully-modeled OS for \texttt{actions\_target} to succeed on a false target. Any information gained about the targets, users, properties of the network and the objectives for the attack can be polluted at every stage, to give the attacker a false sense of success. Even if the attacker learns of deception, untangling which features of the target network are deceptive wastes attacker time and resources, and gives a defensive observer more knowledge about the attacker's response to deception.}

High-interaction decoys and responsive deception tactics have the potential to \nnn{magnify} disruption to the attacker. 
Other taxonomies of deception actions exist to guide the creation of a more robust deception action space~\cite{soule2016}\cite{park2019}.

\bgroup
\def\arraystretch{1.5}
\begin{table}[!htb] \small
	\begin{tabular}{ | p{0.3\columnwidth} | p{0.55\columnwidth} | }
		\hline
		\cellcolor{ltgray}Defensive Actions & \cellcolor{ltgray}Description \\ \hline
		\texttt{do\_nothing} & Null or ``'wait``. The state has not changed, or no action recommended. \\ \hline
		\texttt{monitor\_IDS} & Evaluate network intrusion alerts\\ \hline
		\texttt{monitor\_HBSS} & Evaluate host intrusion alerts\\ \hline
		\cellcolor{ltgray} Deception Action & \cellcolor{ltgray}Description\\ \hline
		\texttt{launch\_decoy} & Creation and deployment of custom decoys. Complementary actions include creating a \texttt{decoy\_user} account or \texttt{plant\_creds} on a real machine to lure attacker to this decoy. \\ \hline
		\texttt{ping\_responder} & Respond to ping request, even if no target exists at that address. \\ \hline
		\texttt{portspoof} & Falsify ports and services on real or decoy hosts. \\ \hline
		\texttt{falsify\_response} & Falsify a success or fail message for a login or information retrieval attempt \\ \hline
		\texttt{traffic\_control} & Change speed of traffic, resulting in a delay or frustration for attacker. \\ \hline
		\texttt{TCP\_reset} & Drop network connection. Delays or deters (redirects) attacks. \\ \hline
		\texttt{create\_user} & Create a false user profile to lure the attacker toward decoy services. \\ \hline
		\texttt{plant\_creds} & Plant fake credentials to waste attacker's time, or working credentials to lure attacker towards a decoy. \\ \hline
	\end{tabular}
	\caption{\textbf{Defender's Action Space:} Simplified list of defensive (observational) and deceptive actions to plant false information and disrupt an attacker's progress.} 
	\label{table:defender_actions}
\end{table}
\egroup

\subsection{State Space}~\label{sub:state}
For an autonomous solution to dynamically determine which deception feature to employ, it will have to calculate the utility of all possible actions given the current state. This state is defined by the actions taken until this point in time, and the features of the current state. The probability of transitioning from one state to the next will drive the type and placement of the deception. These transition probabilities and their calculations over $\{state, action\}$ pairs are expanded upon in Section~\ref{sub:irl}. Different features will hold different values \nnn{during each state} for the attacker, and will drive their actions forward. The defender may not know or share those same values, and it is up to the AI algorithm to determine those values based on prior \nnn{patterns of behavior} and the attacker profile.

\begin{figure}[!h]
	\includegraphics[width=\linewidth]{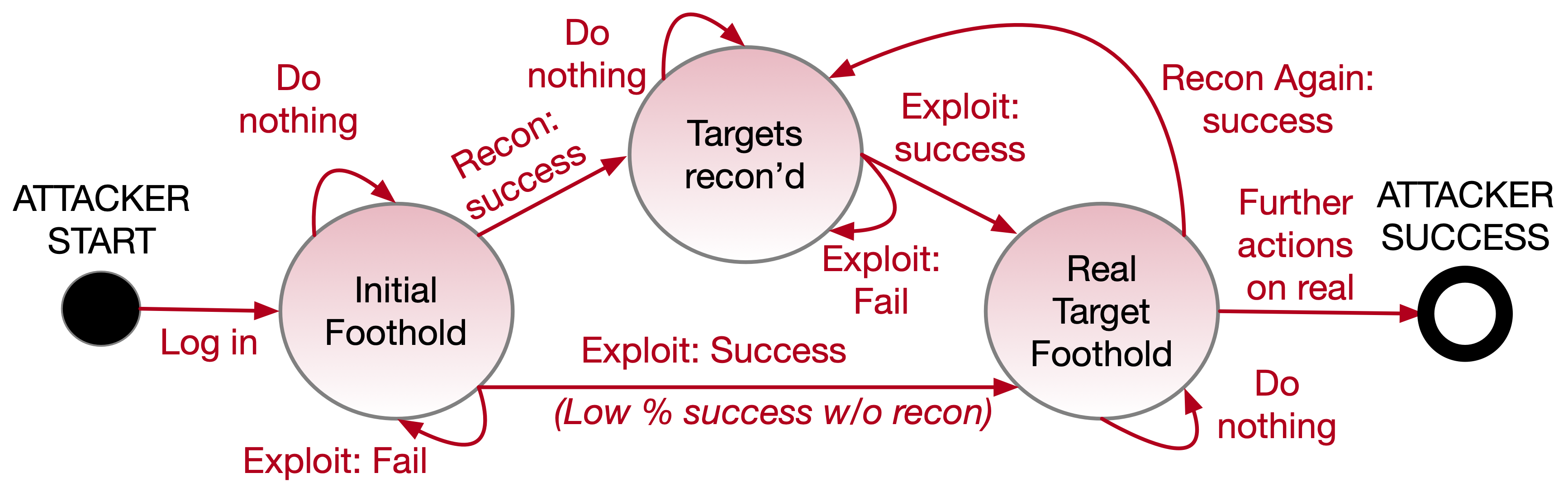}
	\caption{\textbf{Attack State Transition}: Extremely simplified action-transition diagram for a cyber adversary, with no deception present.}
	\label{fig:attacker_state}
\end{figure}

\begin{figure}[!h]
	\includegraphics[width=\linewidth]{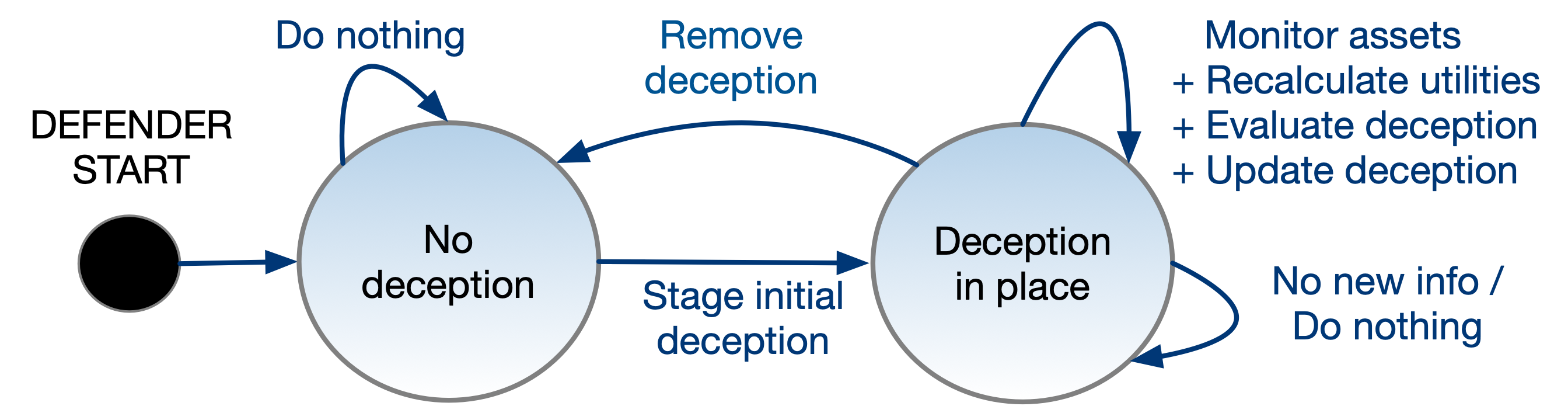}
	\caption{\textbf{Deception State Transition}: Action-transition diagram for a deceptive defender.}
	\label{fig:deception_state}
\end{figure}

A simple example of adversarial state transitions is depicted in Figure~\ref{fig:attacker_state}. Without the presence of deception, given enough tries, the attacker will achieve their objectives. A deceptive defender's simple action transition is depicted in Figure~\ref{fig:deception_state}, however the decisions to refresh the state will depend heavily on observations of the target network and the AI algorithms making deception decisions.

\subsection{Adversary Profiling}~\label{sub:profile}
The observable network features and attacker and defender actions discussed \nnn{previously can} provide \nnn{a stream of} information about \nnn{current and prior} states to an autonomous deception system. 
\nnn{In particular, metadata about features and actions that are definitively associated} with the attacker's activity can provide contextual detail toward profiling inferences about the attacker and their skills, goals, and tactics, which can in turn inform customized deception responses to exploit the attackers feelings and beliefs. 
Some of these metadata data are depicted in Table~\ref{table:profile_known}.

\bgroup
\def\arraystretch{1.5}
\begin{table}[!htb] \small
	\begin{tabular}{ | p{0.3\columnwidth} | p{0.55\columnwidth} | }
		\hline
		\cellcolor{ltgray}\thead[l]{Characteristic} & \cellcolor{ltgray}Description\\ \hline
		Target interactions & The level of interaction, and repeated interactions \\ \hline
		Attack duration & Time since first alert, time to each success/failure \\ \hline
		Stealth & Alert severity and frequency	 \\ \hline
	\end{tabular}
	\caption{\textbf{Attacker Profile:} Metadata about ground truth activities can inform inferences about the attacker's goals and strategies.} 
	\label{table:profile_known}
\end{table}
\egroup

Information which can be inferred about the attacker requires support from prior research and relies heavily on cognitive observations of adversarial activity\cite{Abbasi2000}\cite{gutzwiller_oh_2018}. Data from observable behavior needs to be translated through the lens of foundational research in order to deduce implicit characteristics. For example, patterns of adversarial activity can reveal possible associations to known APTs, as depicted in the ATT\&CK framework~\cite{mitre_attck}. Some inferred characteristics of an attacker are listed in Table~\ref{table:profile_inferred}.

\bgroup
\def\arraystretch{1.5}
\begin{table}[!htb] \small
	\begin{tabular}{ | p{0.3\columnwidth} | p{0.55\columnwidth} | }
		\hline
		\cellcolor{ltgray}Inferred Data & \cellcolor{ltgray}Description\\ \hline
		Sentience & Is the adversary human or automated? (e.g., a bot or scripted attack) 	 \\ \hline
		Number of attackers & How many unique adversaries are present? Are they working collaboratively? Competitively? \\ \hline
		Expertise & Skills, experience, and possibly demographic (e.g. nation state)	 \\ \hline
		Deception-Aware & Is the attacker aware of deception? \\ \hline
		Emotional State & Confusion, Frustration, Self-Doubt \\ \hline
		Goal & Specific target, attack purpose 	 \\ \hline
		Strategy & Tools, techniques, attack patterns	 \\ \hline
		Threat level & Based on expertise and goals, how dangerous is this adversary? \\ \hline
		Fingerprint & Previously seen? APT fingerprinted? 	 \\ \hline
	\end{tabular}
	\caption{\textbf{Inferred Attacker Profile} Information about the attacker that can be determined implicitly via attack patterns and external research.} 
	\label{table:profile_inferred}
\end{table}
\egroup

\nnn{Profiling changes in attack strategy relies far less on inference than measuring an attacker's beliefs or cognitive state in response to deception.} 
While the Tularosa Study demonstrated increased confusion and frustration among participants experiencing deception, the exact reason was not detailed in the findings~\cite{tularosa2019}. 

\nnn{Tularosa participants were surveyed for} cognitive abilities that might be relevant to the \kjf{task}, via well-established surveys and tests to measure overall cognitive ability, fluid intelligence, decision-making confidence, operational memory, and convergent creative thinking. Personality assessments included the Big Five Inventory, the General Decision-Making Style Inventory, the Indecisiveness Scale, and the Need for Cognition~\cite{tularosa2019}. While the answers to these assessments are not something expected to be available on attackers in the wild, they could be combined with existing research on known cyber attacker profiles \nnn{to help construct} attack profiles linked to cyber behavior which can be observed by defenders. 

\section{Reward Modeling with IRL}\label{sub:irl}

\nnn{Deception can be leveraged as a cyber attack deterrent, but optimizing the placement and type of deception requires more than a ``set and forget'' deception policy. This is where autonomous solutions can shine, by strategically taking in data on the state of an attack and learning the attacker's belief in the utility of their own actions to enable predictive and iteratively-optimized deception.}
RL requires a reward function that is able to provide feedback to an agent based on the actions it selects and the states it arrives in, so it can learn the best policy~\cite{sutton_reinforcement_2017}. However, in a cyber scenario, the actions of the attacker cannot be \kjf{directly} controlled, only observed, and the expected utility of responding to those actions relies on the attacker's prior pattern of behavior. Engineering a meaningful reward function for such a complex domain is not a simple task, \nnn{however,} RL can be used to calculate a suitable reward function based on modeled behavior. IRL has \nnn{already demonstrated effectiveness in} capturing expert human route planning~\cite{ziebart2008} and mimicking human fine motor control in robots~\cite{finn2016}. The standard approach to IRL can be difficult to implement when working with static datasets, however, batch IRL with counterfactuals has been used on static healthcare data to understand the decision making of expert doctors. Introducing counterfactuals to batch IRL allows for off-policy evaluations and considers what would happen if different actions were taken at given observation points~\cite{bica2021}. \nnn{Thus, IRL provides potential to reverse engineer rewards from static datasets such as the Tularosa Study, and also has application to learn from cyber defense data on a network in real-time.} 

\nnn{Understanding how humans \kjf{learn} to achieve a goal can help construct the components of an attack learning model.} 
When humans perform \emph{apprenticeship learning}, they do not usually mimic the exact behavior they are trying to learn~\cite{abbeel2004}. Rather, humans observe behavior and presume the intent, then ``copy'' the intention in the sense that their actions accomplish the same goal of the behavior they are trying to learn, but the actions themselves are not necessarily exact replicas of what they have observed~\cite{warneken2006}.
This variability in actions suggests that human behavior is directed by a ``soft optimality'' as opposed to a strictly deterministic, maximal policy for behavior. Probabilistic graphical models can model the soft optimality of human behavior; inference in these models quantifies the probability that an observed trajectory of state and actions pairs was accomplished by a human acting optimally. 
The probability that a human is behaving optimally for a given state-action pair is modeled with a Bernoulli random variable $p(O_t|s_t,a_t)=exp(r(s_t,a_t))$,
where $r(s,a)$ is the reward associated for that state-action pair. 
The most optimal trajectories are the most likely, whereas suboptimal trajectories are exponentially less likely as a function of rewards associated with state action pairs that comprise those trajectories. 

Traditionally, goal directed behavior is modeled by RL, where state space, $s \in S$, and action space, $a \in A$, \nnn{are defined} such that an action $a$, taken in state $s$, may change the current state to a new state, $s'$. These changes may be specified by transition probabilities $p(s'|s,a)$, which give the probability of a transition to state $s'$ from state $s$ when action $a$ is taken in state $s$. These transition probabilities may be given or observed. We also must define a reward function $r(s,a)$. The goal of RL is to learn an optimal policy $\pi^*(a|s)$, which specifies what action to take in each state so as to maximize the expected value of the reward.

In general, a reward function may be difficult to specify; however, data associated with performing a \nnn{given action} may successfully provide the opportunity to use IRL techniques to capture that reward function, \nnn{which can} then be used to re-optimize the reward function with forward RL algorithms. For example, in the Tularosa Study the reward function that is driving the experts' behavior \nnn{is unknown}, but \nnn{data provides} states, actions, and sample trajectories $\tau_i$ \nnn{observed from participant data}. \nnn{Without knowing the optimal policy, we} may assume that we have observed a close approximation of the optimal policy \nnn{through} observing the behavior of many experts.
\nnn{From this,} the reward function is modeled as $r_\psi(s,a)$, where $\psi$ is a vector that parameterizes the reward that the expert's policy optimized to produce the observed set of trajectories, ${\tau_i}$. The reward function may be parameterized in several ways, for example, as a linear reward function as a weighted sum of features associated with a state-action pair, 
$r_\psi(s,a)=\sum_i\psi_if_i(s,a)$. Alternatively, a neural network with parameter vector $\psi$ may be used to specify the reward function. \nnn{IRL holds a strong potential to} capture the reward function from the observed trajectories of behavior performed by an expert. Instead of \nnn{trying to learn} the probability of an attack action given a reward, which amounts to inference in the graphical probabilistic model, IRL uses the same graphical model to learn the reward function. By observing the given trajectories, we can learn the parameters of the rewards so that the likelihood of the observed trajectories in the graphical model are maximized~\cite{levine2018}. 

The data from the Tularosa Study provides observations of human experts performing a cyber-attack task. Often the reward functions use for cyber security \kjf{domains} can seem arbitrary and it is unclear as to whether they will transfer to actual human cyber-attack behavior. If instead, IRL can be used to learn an attacker's reward function from experimental (e.g., Tularosa), capture-the-flag (CTF), or real-world data, then this could: 1) be applied to an attacking RL agent that will help train an autonomous defender and 2) be used to infer the reward function of the defender (assuming zero-sum) and used directly to train an autonomous \nnn{deceptive} defender.  \nnn{Further} work \nnn{can also include} new HSR experimentation focused on defender actions used in conjunction with IRL to measure the defender's reward function directly. 

\section{Discussion}~\label{sec:conclusion}
While the current conversation around intelligent cyber defense is often centered around the discovery of malicious or suspicious activity, this is primarily a reactive measure.  Armed with defensive deception along with visibility into network and host data, a defender can \emph{proactively} detect, monitor, and neutralize attacks at every stage. There is an increasing need for AI applied to cyber defense to handle timely and effective deployment of reactive and proactive responses, as well as to customize the deceptive responses to the needs of the defender and the learned profiles of the attacker. 

In this paper we present a simplified set of features available to a cyber defender, which include deceptive actions, each of which can learn to adapt in real time to a malicious actor. Examples of the effectiveness of deception as presented in the Tularosa Study serve to showcase that these data can be measured to evaluate---or support the \kjf{use of}---this type of deception. Attacker profiling is another layer of continually-learned metadata collected by the defender's own sensors, and can be used to inform type and placement of deception to disrupt an attacker's decision points and pollute their forward progress. A strong reward function is the backbone of any autonomous solution. We provide an example of how learning from data inputs via IRL can reveal the attacker's own reward function, which can help inform the attacker profile, and the defender's deceptive response.

 Through this research we aim to instill an understanding of how and why insights from real-world and HSR cyber attacker data are necessary components to guide the advancement of intelligent defense systems. \nnn{Autonomous cyber deception systems can leverage HSR and attack data to learn adversarial reward functions, and exploit this knowledge through optimized deception placement} \kjf{and configuration.} Future work entails \kjf{implementing IRL algorithms on Tularosa data, as well as} utilizing real-world data or \kjf{designing} new HSR to investigate \emph{defensive} decision points to infer how experts counter and remediate attacks. 

%
\section*{Acknowledgments}
%
This work was partially funded by Cyber Technologies, C5ISREW Directorate, Office of the Under Secretary of Defense Research and Engineering as well as the Laboratory for Advanced Cybersecurity.

\bibliographystyle{plain}
\small
\bibliography{bibliography5.bib}
\normalsize

\end{document}